\documentclass[%
reprint,
amsmath,amssymb,
aps,
prl,
floatfix,
]{revtex4-2}

\usepackage{graphicx}
\usepackage{dcolumn}
\usepackage{bm}
\usepackage{layouts}
\usepackage{verbatim}
\usepackage[dvipsnames]{xcolor}
\usepackage{siunitx}
\usepackage{hyperref}
\usepackage{placeins}
\usepackage{chemformula}
\usepackage{booktabs}
\usepackage[autostyle]{csquotes}

\graphicspath{{figures/}} 

\hypersetup{colorlinks,
    linkcolor={red!80!black},
    citecolor={blue!80!black},
    urlcolor={blue!80!black}}

\DeclareRobustCommand{\rchi}{{\mathpalette\irchi\relax}}
\newcommand{\irchi}[2]{\raisebox{\depth}{$#1\chi$}} 

\DeclareSIUnit{\belmilliwatt}{Bm}
\DeclareSIUnit{\dBm}{\deci\belmilliwatt}

\begin{document}
\preprint{}

\title{Inductive magnon noise spectroscopy}

\author{Luise Siegl$^1$}
\author{Richard Schlitz$^1$}
\author{Jamal Ben Youssef$^2$}
\author{Christian Runge$^1$}
\author{Akashdeep Kamra$^3$}
\author{William Legrand$^4$}
\author{Hans Huebl$^{5,6,7}$}
\author{Michaela Lammel$^1$}
\author{Sebastian T. B. Goennenwein$^1$}

\affiliation{$^1$Department of Physics, University of Konstanz, 78457 Konstanz, Germany}
\affiliation{$^2$LabSTICC, CNRS, Université de Bretagne Occidentale, 29238 Brest, France}
\affiliation{$^3$Condensed Matter Physics Center (IFIMAC) and Departamento de F\'{i}sica Te\'{o}rica de la Materia Condensada, Universidad Aut\'{o}noma de Madrid, E-28049 Madrid, Spain}
\affiliation{$^4$Université Grenoble Alpes, CNRS, Institut Néel, 38042 Grenoble, France}
\affiliation{$^5$Walther-Meißner-Institut, Bayerische Akademie der Wissenschaften, 85748 Garching, Germany}
\affiliation{$^6$TUM School of Natural Sciences, Technische Universität München, 85748 Garching, Germany}
\affiliation{$^7$Munich Center for Quantum Science and Technology (MCQST), 80799 München, Germany}

\date{\today}

\begin{abstract}
	State tomography allows to characterize quantum states, and was recently applied to reveal the dynamic magnetization state of a parametrically driven magnet.
	The identification of non-classical states, such as squeezed states, relies on a careful analysis of their emission and their distinction from thermal and vacuum fluctuations.
	A technique allowing to detect equilibrium magnetization fluctuations is a crucial first step in this regard. 
	In this Letter, we show that inductive magnon noise spectroscopy (iMNS) allows to characterize the thermal magnetization fluctuations of a ferromagnetic thin film in a broadband coplanar waveguide-based scheme.
	Relative to a cold microwave background, the microwaves emitted by the equilibrium magnetization fluctuations can be detected via spectrum analysis.
	We provide a comprehensive picture of our microwave system by quantitatively modeling its response, including the thermalizing influence of the cables.
	The model allows for direct comparison to low-power broadband ferromagnetic resonance measurements with excellent agreement, corroborating the equilibrium character of the iMNS measurement by probing the linear response of the equilibrium state.
	Our work thus demonstrates broadband access to the equilibrium properties of magnetization fluctuations using a purely inductive approach. 
\end{abstract}

\maketitle

Electrical noise spectroscopy is a generic and powerful approach, routinely applied to the characterization of classical and quantum systems~\cite{DePicciotto1997,Saminadayar1997,clerk_introduction_2010}.
As a particular application of noise spectroscopy, state tomography is used for the experimental investigation, identification, and understanding of the quantum state of a system \cite{Altepeter2005,Hioki2021}. 
In the microwave domain, tomography techniques have been perfected, e.g., to characterize the photon statistics of propagating thermal microwaves \cite{Goetz2017,Menzel2010} or to study the displacement of propagating squeezed microwave states \cite{Fedorov2016}. 

In turn, quantum states of magnetic systems and in particular non-classical magnon states, such as squeezed magnon modes \cite{Kamra2016a, Kamra2016b, Haghshenasfard2017, Li2019, Kostylev2019, Kamra2019, Kamra2020a, Elyasi2020, yuan_quantum_2022} or magnons entangled with other (quasi-) particles like phonons or photons \cite{Li2018, Li2019a, Zhang2019a, lachance-quirion_hybrid_2019, Yuan2020, Elyasi2020, Zou2020, Mousolou2021, Wuhrer2022}, are of interest, e.g., for quantum enhanced sensing applications. 
Most previous approaches towards their characterization are qubit-based and thus require experiments at low temperatures \cite{Tabuchi-Nakamura-2015, lachance-quirion_entanglement-based_2020, wolski_dissipation-based_2020, yuan_quantum_2022, Römling2023}, since most qubits operate only at low temperatures and the conventional magnon modes in a magnet are bosons with a spectrum of eigenenergies reaching down to almost zero energy \cite{Stancil-spin-waves-book}.

The development of magnetization state tomography (MST) based on electrical measurements is still in its infancy \cite{Hioki2021,Sharma2023}.
First works demonstrated a mixed phase state that arises when pairs of magnons are parametrically excited by microwave radiation \cite{Hioki2021, Makiuchi2024}.
When driven even further out of equilibrium, magnetic systems moreover can reach states with characteristics similar to superconductors (referred to as magnon Bose-Einstein condensates) even at room temperature~\cite{demokritov_boseeinstein_2006,pirro_advances_2021,divinskiy_evidence_2021,mohseni_classical_2022}.
To establish potential quantum effects in such condensates, measurements of the magnetic driven state at room temperature as well as its equilibrium properties are essential as a reference.
However, the microwave electronics typically used to measure the magnetic state are subject to electronic Johnson-Nyquist (JN) 
noise with an amplitude comparable to that of the magnetic fluctuations \cite{Johnson1928, Nyquist1928}. 
Consequently, it is difficult to distinguish the steady-state thermal microwave emissions that originate from a magnetic system coupled to the electric circuit from the inherent electronic noise of the circuit itself.
Fortunately, the JN noise can be modulated by lowering the temperature of the electronic part of the system \cite{Dicke1946}.
By operating the detection system at a lower temperature compared to the magnetic system, I.~A.~Deryugin and N.~I.~Lyashenko were able to detect the fluctuations of the collective magnon (Kittel) mode in bulk garnet crystals about 60 years ago in narrow-band microwave-based experiments \cite{Deryugin1963}, which have not been repeated since.
In addition to inferring the equilibrium magnetic state defined by the magnon noise temperature, this approach also appears attractive for magnon state preparation, e.g., by selectively cooling the magnonic state with respect to the lattice temperature of the solid \cite{Fassioli2024}.

\begin{figure}
	\includegraphics{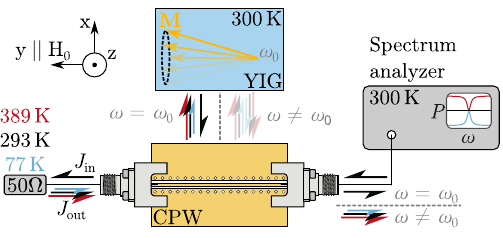}%
	\caption{\label{fig1:concept}
	A schematic illustration of the experimental setup consisting of a spectrum analyzer, a CPW and a termination ($\SI{50}{\ohm}$). 
	Increasing ($\SI{389}{\kelvin}$) or decreasing ($\SI{77}{\kelvin}$) the termination's temperature from ambient temperature ($\SI{293}{\kelvin}$) leads to a non-equilibrium state of the system and a net energy flux (red and blue arrows) arises. 
	In thermal equilibrium all system components have the same temperature, leading to a net zero energy flux (black arrows). 
	The YIG thin film resides at the ambient temperature of $\SI{293}{\kelvin}$ and, thus, always emits and absorbs energy in the form of fluctuations with an enlarged spectral weight at the FMR frequency $\omega_0$. 
	Thus, depending on the temperature of the termination, its emitted energy flux $J_{\rm{out}}$ can be larger (red arrow) or smaller (blue arrow) than in equilibrium. 
	}
\end{figure}

In this Letter, we demonstrate inductive magnon noise spectroscopy (iMNS), a conceptually simple, broadband, microwave-based technique allowing to characterize thermal magnetization precession fluctuations in a magnetic sample.
We observe an enhancement in the noise power spectral density close to the ferromagnetic resonance (FMR) frequency $\omega_0$ in a ferromagnetic thin film at room temperature resulting from fluctuations of the magnetization, corroborating the results put forward by Deryugin and Lyashenko \cite{Deryugin1963}.
In contrast to Ref.~\cite{Deryugin1963}, our broadband experiments rely on varying the temperature of the $\SI{50}{\ohm}$ termination of the microwave circuit (see Fig.~\ref{fig1:concept}).
This modulates the electronic thermal noise, such that the thermal microwaves emitted from the magnetic thin film become accessible.
In other words, we measure the magnetization fluctuations in equilibrium (driven by the thermal bath of the lattice) by connecting the magnet inductively to a microwave environment at a different temperature, which allows extracting information about the thermal state of the ferromagnetic thin film. 
We quantitatively assess our observation using the equipartition theorem and further show that thermal microwaves emitted from the $\SI{50}{\ohm}$ termination can be used for noise-driven FMR experiments. 
We critically compare the observed line shape and width with conventional, driven, broadband FMR measurements to corroborate the notion that iMNS indeed probes the linear response of the equilibrium state.

In our experiments the magnetic sample is an yttrium iron garnet (YIG) film grown by liquid phase epitaxy (LPE) on a (111)-oriented gadolinium gallium garnet substrate. 
The YIG layer has a thickness of $\SI{74}{\nano\meter}$, a low Gilbert damping $\alpha = \num{2.83(0.02)e-4}$ and an inhomogenious contribution to the line broadening $\Delta f_0 = \SI{4.60(0.07)}{\mega\hertz}$. 
The sample is glued on a coplanar waveguide (CPW) with a center conductor width of $\SI{90}{\micro\meter}$. 
A static external magnetic field $\mu_0 H_0$ is applied along the CPW in the plane of the YIG layer. 
One port of the CPW is passively terminated with a $\SI{50}{\ohm}$ termination resistor (in the following called termination), resulting in an impedance matched electrical circuit (see Fig.~\ref{fig1:concept}). 
The termination can be heated or cooled independently of the rest of the electric circuit. 
The noise power density spectrum $S_{\rm{n}}(\omega, \mu_0 H_0)$ is measured as a function of frequency $\omega$ for a series of static magnetic field strengths using a  spectrum analyzer connected to the other port of the CPW as shown in Fig.~\ref{fig1:concept}. 
The data hereby are taken using the fast Fourier transform (FFT) mode and the root mean square detector of the spectrum analyzer, with a resolution bandwidth of $\Delta f = \SI{1}{\kilo\hertz}$ and the internal preamplifier set to a gain of $\SI{30}{\decibel}$. 
Additionally, the spectrum analyzer's internal noise cancellation function is used to cancel its inherent noise \cite{RS_manual2024}. 
Furthermore, the microwave circuit is carefully designed to minimize parasitic signals such as local oscillator leakage from the spectrum analyzer. 
To remove ripples resulting from remaining parasitic contributions, a background subtraction method is used (see Supplemental Material, SM~\cite{SM} for details). 
To characterize the sample properties, we use conventional, driven, broadband FMR measurements.
To that end, we replace the spectrum analyzer and the termination with a two-port vector network analyzer (VNA), i.e., we connect the VNA to both ports of the CPW. 
Thus, the FMR of the YIG layer can be detected as change of the complex scattering parameter $S_{21}(\omega, \mu_0 H_0)$ at resonance conditions.

\begin{figure*}
	\includegraphics[width=\textwidth]{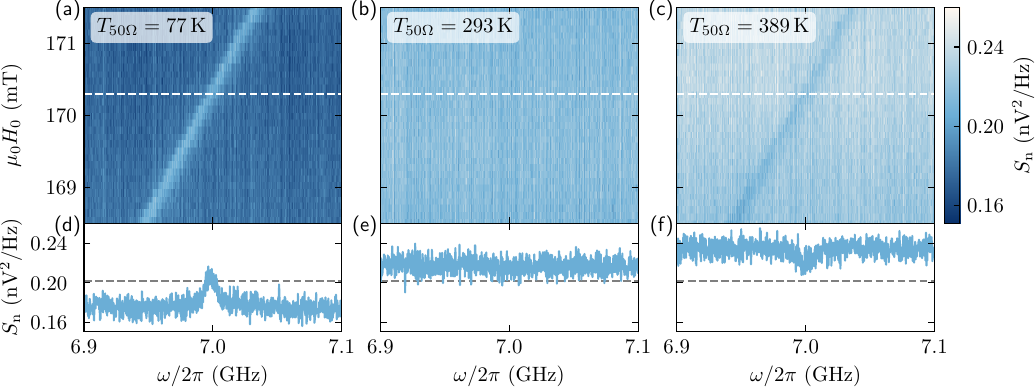}%
	\caption{\label{fig2:temp_dependence}
	Changing the temperature of the termination $T_{\rm{50\Omega}}$ allows to observe different phenomena. 
	Either magnon noise is detected via the corresponding microwave fluctuations (a) and (d), a constant microwave background is seen (b) and (e), or we find noise driven FMR (c) and (f). 
	(a)--(c) False-color plots of the noise power spectral density $S_{\rm{n}}$ as a function of microwave frequency $\omega$ and static magnetic field $\mu_0H_0$ are shown for the different termination temperatures $T_{\rm{50\Omega}}$. 
	(d--f) Line cuts at fixed magnetic fields ($\mu_0H_0=\SI{170.2}{\milli\tesla}$) corresponding to the panel above. 
	The dashed gray line indicates the JN noise at $T=\SI{293}{\kelvin}$.
	The noise emitted (d) and noise driven (f) FMR mode appear as a peak or dip on the background noise level, respectively.
	(e) With all setup components at ambient temperature, only a constant noise power level is visible. 
	}
\end{figure*}

We now qualitatively describe the operational principle of iMNS. 
In the presence of a thermal gradient in the microwave circuit, beyond diffusive heat transport, an energy flow occurs along the transmission line in the form of an imbalance of microwave electromagnetic radiated noise \cite{Fourier1878}. 
Thus, no net energy flux should be present if all components of the setup are in thermal equilibrium. 
Then, the spectrum analyzer will detect microwave radiation with a power density corresponding to the thermal energy, which is equivalent to the JN noise \cite{Johnson1928,Nyquist1928,Dicke1946} of a $\SI{50}{\ohm}$ resistor at temperature $T$, as our microwave setup is designed with a $\SI{50}{\ohm}$ impedance (see black arrows in Fig.~\ref{fig1:concept}).
In turn, if the termination is cooled well below ambient temperature, its microwave emission is reduced.
The magnetization precession induced by thermal fluctuations now results in a net energy flow into the microwave circuit with a nonuniform spectral distribution and therefore can be distinguished from the broadband thermal background by the spectrum analyzer (see blue arrows in Fig.~\ref{fig1:concept}). 
Thus, the noise emission of the equilibrium magnetization fluctuations can be detected. 
A similar, but not broadband, approach was used by Deryugin and Lyashenko, who used horn antennas pointed at the sky to obtain a cold background over which the thermal emission from the magnetization fluctuations of a magnet can be detected~\cite{Deryugin1963}. 
Our approach does not only allow to cool but also to heat the background given by the electric circuit.
If the termination is heated, it emits a frequency independent (white) JN noise of larger thermal magnitude as compared to the sample at room temperature. 
This contribution then drives the magnetic fluctuations of the YIG layer, which will result in a net absorption of power close to the FMR frequency $\omega_0$, as in conventional FMR measurements (see red arrows in Fig.~\ref{fig1:concept}).
We thus term this situation noise driven FMR.

Figure~\ref{fig2:temp_dependence} shows the noise power spectral density $S_{\rm{n}}$ detected by the spectrum analyzer as a function of static magnetic field $\mu_0 H_0$ and frequency $\omega$ for three distinct temperatures (from left to right) of the termination. 
In the top panels (a)--(c) the full spectrum is shown as a false-color plot and in the bottom panels (d)--(f) the line cuts for a magnetic field of $\mu_0 H_0 = \SI{170.2}{\milli\tesla}$ are shown. 
As per the above discussion, when the termination is at ambient temperature $T_{\rm{50\Omega}}=\SI{293}{\kelvin}$ (c.f. Fig.~\ref{fig2:temp_dependence} (b) and (e)), the noise $S_{\rm{n}}(\omega)$ only shows a constant background level.
The dashed gray line (panels (d)--(f)) indicates the JN noise $S_{\rm{JN}} = k_{\rm{B}} T R$ \cite{Johnson1928,Nyquist1928} with $R = \SI{50}{\ohm}$ and $T=\SI{293}{\kelvin}$.
The noise power observed in experiment is systematically about $\SI{10}{\percent}$ larger, which will be discussed later.
Upon cooling the termination, the background signal, naively dominated by the JN noise of the termination is reduced as expected.
Additionally, we observe an enhancement of the detected microwave power at the frequency $\omega_0$ of the Kittel mode, consistent with the corresponding $\omega$ and $\mu_0 H_0$, measured in a conventional FMR experiment.
In turn, when the termination is heated (c.f. Fig.~\ref{fig2:temp_dependence} (c) and (f)), the observed noise power is increased and exhibits a dip at $\omega_0$, as the YIG layer absorbs the excess of thermal microwaves photons, consistent with the noise driven FMR introduced above.
The observed features, thus, confirm the qualitative operation principle introduced above and showcase that the microwave emission from magnetization fluctuations in thermal equilibrium can be detected in our measurement scheme.

\begin{figure}[tb]
	\includegraphics[width=\linewidth]{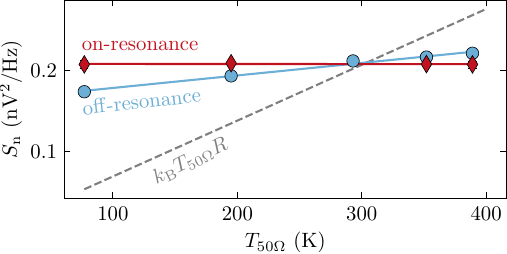}%
	\caption{
		Dependence of the noise power spectral density $S_{\rm{n}}$ on the termination temperature $T_{\rm{50\Omega}}$. 
		The common expression of the JN noise (dashed line), strongly deviates from the experimental data (blue circles and red diamonds). 
		A modified expression, see Eq.~(\ref{eq:background_noise}), describing the noise power spectral density of the electronic circuit including noise coupling into the transmission line, is used for fitting the on- and off-resonance data (red and blue line). 
	\label{fig3:background}}
\end{figure}

Quantitatively, however, we find a large discrepancy between the measured off-resonance noise power and the expected JN noise of the termination ($S_{\rm{n}}(\omega\neq\omega_0) \neq S_{\rm{JN}}$) as shown in Fig.~\ref{fig3:background}. 
The off- and on-resonance noise power spectral density $S_{\rm{n}}$ shown in Figure~\ref{fig3:background} for different termination temperatures $T_{\rm{50\Omega}}$ are obtained by fitting high resolution (1000 averages) iMNS measurements with a Lorentzian function and a constant background (see SM~\cite{SM}). 
We ascribe the difference between the expected and observed noise to additional noise introduced by the attenuation of the transmission lines (microwave coaxial cables, connectors, etc)~\cite{pozar_microwave_2012}. 
For a more rigorous description of the microwave circuit we must therefore consider the influence of the coaxial cables and the CPW. 
Their finite losses lead to a below unity power transmission $\mathcal{T} = P_\mathrm{out}/P_\mathrm{in} < 1$ (defined as the power ratio of the output and input power) through the assembly. 
Therefore, the microwave circuit attenuates the JN noise emitted by the termination by $\mathcal{T}$. 
Additionally, the losses leading to $\mathcal{T} < 1$ cause the environment to couple into the transmission line with the environmental temperature (room temperature in our case), resulting in additional noise with a power spectral density of $S_{\rm{env}} = k_{\rm{B}} T_{\rm{env}} R$ at a temperature of $T_{\rm{env}} = \SI{293}{\kelvin}$ and resistance $R = \SI{50}{\ohm}$.
This added noise power contributes with $(1-\mathcal{T})$ since larger losses in the transmission line lead to a more efficient thermalization of the microwaves, parameterized by a smaller $\mathcal{T}$ and thus enhanced contribution of $S_{\rm{env}}$~\cite{blais_circuit_2021}. 
Accounting for both contributions, the off-resonance level of the noise power spectral density can be described by
\begin{equation}\label{eq:background_noise}
    S_{\rm{n}}(\omega\neq\omega_0) = \mathcal{T} k_{\rm{B}} T_{\rm{50\Omega}} R + (1-\mathcal{T}) S_{\rm{env}} + S_{\rm{off}} , 
\end{equation}
where $S_{\rm{off}}$ is an additional constant offset caused, e.g., by the imperfect noise cancelling of the spectrum analyzer and the background subtraction.
Using Eq.~(\ref{eq:background_noise}) and $R = \SI{50}{\ohm}$, we find $\mathcal{T} = \num{0.22(0.01)}~\widehat{=}~\SI{-6.5(0.3)}{\decibel}$ and $S_{\rm{off}} = \SI{0.006(0.001)}{\nano\volt\squared\per\hertz}$. 
In turn, the on-resonance noise power spectral density is almost independent of temperature, showing that the noise originating from the termination is mostly attenuated due to the strong coupling between the magnetic sample and the CPW. 

\begin{figure}[tb]
	\includegraphics[width=\linewidth]{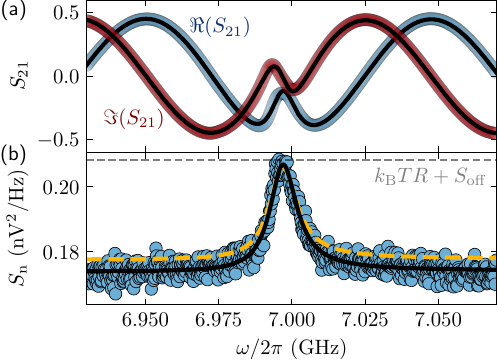}%
	\caption{\label{fig4}
	(a) VNA measurement of the FMR at a power of $\SI{-30}{\dBm}$. 
	Fitting of the real and imaginary part of $S_{21}(\omega)$ with the model in Eq.~(\ref{eq:S21_model}) yields the solid lines.
	(b) MNS measurement with high resolution (1000 averages) and $T_{\rm{50\Omega}}=\SI{77}{\kelvin}$. 
	Using the same parameters from the $S_{21}$ fit shown in panel (a) and $S_{\rm{off}} = \SI{0.006}{\nano\volt\squared\per\hertz}$, we obtain the dashed yellow line. 
	The black line is a fit using the model in Eq.~(\ref{eq:Noise_model}) and the JN noise at $T=\SI{293}{\kelvin}$ plus the constant offset $S_{\rm{off}}$ is represented by the dashed gray line.
	}
\end{figure}

To connect our iMNS measurements to the established FMR models, we elaborate the complex scattering parameter $S_{21}$ and its relation to the noise data using our model (see Eq.~\ref{eq:Noise_model}).
A ferromagnetic sample couples to the external electric circuit according to \cite{Bilzer2007,Probst2015, Schlitz2022}
\begin{equation}\label{eq:S21_model}
    S_{21}(\omega) = A \exp(i\beta) \exp(-i\tau\omega) [1 - \Delta S_{21}(\omega)] ,
\end{equation}
where $A$ is an amplitude attenuation factor, $\beta$ a phase shift and $\tau$ the electrical delay time, all related to the microwave circuit itself. 
The change in $S_{21}(\omega)$ due to the FMR response of the sample is described in a linear perturbation approach by $\Delta S_{21}(\omega) = i\eta \rchi(\omega) \exp(i \phi)$ \cite{Schoen2015}, with dimensionless coupling strength $\eta$ to the microwave circuit, the complex microwave susceptibility $\rchi(\omega)$ \cite{Kalarickal2006} and phase $\phi\neq0$ appearing when the sample causes the CPW to lose its $\SI{50}{\ohm}$ impedance matching. 
We obtain the frequency dependent noise power spectral density from Eqs.~(\ref{eq:background_noise}) and (\ref{eq:S21_model}), using the power transmission $\mathcal{T}\rightarrow$ $|S_{21}(\omega)|^2$ of the circuit, which yields
\begin{eqnarray}\label{eq:Noise_model}
    S_{\rm{n}}(\omega) = &&|A|^2 |1 - \Delta S_{21}(\omega)|^2 k_{\rm{B}} T_{\rm{50\Omega}} R \nonumber\\
    &&+ (1-|A|^2|1 - \Delta S_{21}(\omega)|^2) S_{\rm{env}} + S_{\rm{off}} ,
\end{eqnarray}
where $|A|^2$ corresponds to $\mathcal{T}$ in Eq.~(\ref{eq:background_noise}). 

In Fig.~\ref{fig4} we compare an iMNS measurement to the complex FMR transmission coefficient $S_{21}(\omega)$, both at $\mu_0 H_0 = \SI{170.2}{\milli\tesla}$.
The overall oscillation of $S_{21}(\omega)$ is due to the electrical delay of the microwave circuit. 
We show the real and imaginary part of the $S_{21}$ spectrum (blue and red circles) for a power of the VNA $P_{\rm{VNA}}=\SI{-30}{\dBm}$ in panel (a).
We fit the experimental data using Eq.~(\ref{eq:S21_model}) which yields the black lines (for further details please refer to the SM~\cite{SM}).
In panel (b), we show $S_{n}(\omega)$ and the noise determined by Eq.~(\ref{eq:Noise_model}) (dashed yellow line) with the same parameters as for the $S_{21}$ fit and $S_\mathrm{off} = \SI{0.006}{\nano\volt\squared\per\hertz}$, in excellent agreement with the noise measurement. 
Further fitting the noise data with Eq.~(\ref{eq:Noise_model}) (black line) yields very similar parameters, as detailed in the SM~\cite{SM}.
The excellent agreement between the two measurement approaches validates our model and further corroborates the equilibrium character of the iMNS measurement. 
Please note that the visibility of the ferromagnetic peak in the noise power spectral density is maximized with $\mathcal{T} \eqsim 1$ and a large $|\eta \rchi(\omega)|$ at resonance, which expresses the attenuation due to the resonant ferromagnetic sample. 
For our sample $|\eta \rchi(\omega)| \approx \num{0.8}$ at resonance, which provides a strong signal while the form of $\Delta S_{21}(\omega)$ given above remains a reasonable approximation.

In summary, we have established inductive magnon noise spectroscopy (iMNS) as a conceptually simple experimental tool for probing resonant magnetic noise properties in equilibrium.
In the iMNS experiments, we use a broadband coplanar waveguide-based scheme and a variable-temperature microwave background to discern the microwaves emitted by the room temperature fluctuations of the magnetization in a YIG thin film from thermal noise emitted by the termination.
Accounting for finite attenuation in the microwave cabling allows to model the microwave circuit and obtain excellent agreement with the data. 
Using a purely inductive approach, our work demonstrates a broadband access to the equilibrium properties of magnetization fluctuations, which is a first step towards the detection of quantum states of magnetization by magnetic state tomography.

\FloatBarrier

\begin{acknowledgments}
	We acknowledge financial support from the Deutsche Forschungsgemeinschaft (DFG, German Research Foundation) via the SFB 1432 – Project-ID 425217212. 
	H.H. acknowledges financial support by the Deutsche Forschungsgemeinschaft (DFG, German Research Foundation) via Germany’s Excellence Strategy EXC-2111-390814868 and the Transregio TRR 360 – 492547816.
	A.K. acknowledges financial support from the Spanish Ministry for Science and Innovation -- AEI Grant CEX2018-000805-M (through the ``Maria de Maeztu'' Programme for Units of Excellence in R\&D) and grant RYC2021-031063-I.
	The data shown in the figures will be made openly available and linked to the article with a unique DOI once the article is published.
\end{acknowledgments}

\FloatBarrier

\bibliography{references}
\end{document}


\title{Supplemental Material: Inductive magnon noise spectroscopy}

\author{Luise Siegl$^1$}
\author{Richard Schlitz$^1$}
\author{Jamal Ben Youssef$^2$}
\author{Christian Runge$^1$}
\author{Akashdeep Kamra$^3$}
\author{William Legrand$^4$}
\author{Hans Huebl$^{5,6,7}$}
\author{Michaela Lammel$^1$}
\author{Sebastian T. B. Goennenwein$^1$}

\affiliation{$^1$Department of Physics, University of Konstanz, 78457 Konstanz, Germany}
\affiliation{$^2$LabSTICC, CNRS, Université de Bretagne Occidentale, 29238 Brest, France}
\affiliation{$^3$Condensed Matter Physics Center (IFIMAC) and Departamento de F\'{i}sica Te\'{o}rica de la Materia Condensada, Universidad Aut\'{o}noma de Madrid, E-28049 Madrid, Spain}
\affiliation{$^4$Université Grenoble Alpes, CNRS, Institut Néel, 38042 Grenoble, France}
\affiliation{$^5$Walther-Meißner-Institut, Bayerische Akademie der Wissenschaften, 85748 Garching, Germany}
\affiliation{$^6$TUM School of Natural Sciences, Technische Universität München, 85748 Garching, Germany}
\affiliation{$^7$Munich Center for Quantum Science and Technology (MCQST), 80799 München, Germany}

\maketitle

\section{Background Subtraction Method}
\begin{wrapfigure}{r}{0.5\textwidth}
    \includegraphics[width=0.99\linewidth]{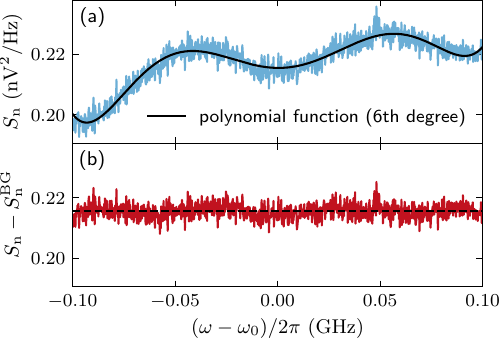}%
    \caption{\label{SI_fig1}
    (a) Polynomial function of degree 6 (solid black line) of the high resolution (1000 averages) data at $T_{\rm{50\Omega}}=\SI{293}{\kelvin}$ and $\omega_0/2\pi = \SI{7}{\giga\hertz}$.
    (b) The polynomial fit is subtracted from the data, excluding the constant offset $c_0$ (dashed line), by Eq.~\ref{SI_eq:background_func}.}
\end{wrapfigure}

Due to parasitic signals, the measured noise power spectral density exhibits finite background ripples. 
To eliminate these ripples while maintaining the constant noise level, we apply a background subtraction method.
We fit the high-resolution (1000 averages) data at $T_{\rm{50\Omega}}=\SI{293}{\kelvin}$ using a polynomial function of degree 6 $\left( \sum_{k=1}^{6}x^k c_k + c_0 \right)$, where $c_0$ is a constant offset (see Fig.~\ref{SI_fig1}(a)).
Since all parts of the setup are in equilibrium at this temperature, the magnetic sample does not contribute. 
In the following, we subtract the noise background
\begin{equation}\label{SI_eq:background_func}
    S_{\rm{n}}^{\rm{BG}} = \sum_{k=1}^{6} x^k c_k , 
\end{equation}
where $c_k$ with $k\in\{1,...,6\}$ are the fit parameters, from the iMNS data taken at different temperatures. 

We choose the x-axis offset ($x=0$) as $\omega-\omega_0$, where $\omega_0/2\pi = \SI{7}{\giga\hertz}$ represents the resonance frequency.
Note that we intentionally exclude the constant offset $c_0$ from the subtraction to maintain the background level of the noise power spectral densities. 
Please note, however, that this value of the background level still might include an error of up to $\SI{10}{\percent}$ from the amplitude of the residual ripple.
Figure~\ref{SI_fig1}(a) shows the high resolution measurement at $T_{\rm{50\Omega}}=\SI{293}{\kelvin}$ and the polynomial fit (including $c_0$) as the solid black line.
In panel (b) the corrected data of the measurement are shown, where Eq.~\ref{SI_eq:background_func} is subtracted from the data.
The dashed line shows the fit parameter of the constant offset $c_0$.

\FloatBarrier

\section{Fitting the high resolution measurements}
\begin{figure}[h]
    \includegraphics[width=0.99\linewidth]{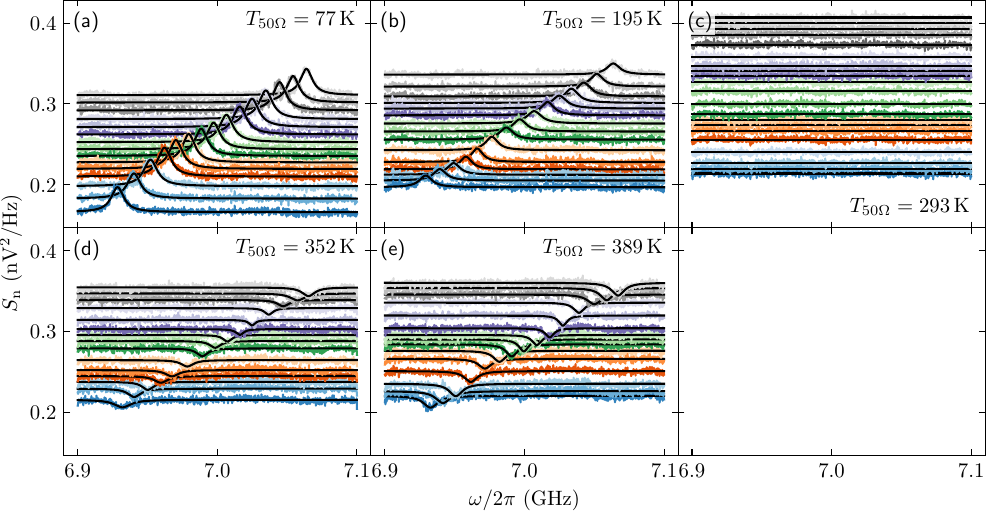}%
    \caption{\label{SI_fig2}
    Noise power spectral density $S_{\rm{n}}$ as a function of frequency $\omega/2\pi$ at different magnetic fields. 
    (a)--(b) and (d)--(e) The data is fitted with a Lorentzian function plus a constant offset, where the black lines indicate the fits. 
    (c) At room temperature the noise signal vanishes, thus, we use only a constant model for fitting. 
    Starting from the dark blue data, all magnetic field cuts are separated by $\SI{0.01}{\nano\volt\squared\per\hertz}$ relative to each other.}
\end{figure}

To evaluate the on- and off-resonance values of the high resolution (1000 averages) iMNS data we fit the measurements from approximately $\SI{168}{\milli\tesla}$ to  $\SI{173}{\milli\tesla}$ with a Lorentzian function plus a constant offset (see Fig.~\ref{SI_fig2} (a)--(b) and (d)--(e)). 
The measurement at room temperature $T_{\rm{50\Omega}}=\SI{293}{\kelvin}$ (cf. Fig.~\ref{SI_fig2} (c)) is fitted with a constant model only. 
From all fits of the corresponding termination temperature we calculate the arithmetic mean and the standard deviation of the constant offset and the noise power at resonance. 
Figure~\ref{SI_fig2} shows the noise power spectral density $S_{\rm{n}}$ depending on frequency $\omega/2\pi$ at different magnetic fields, where the black lines represent the corresponding fits. 
For better visibility, the measurements are shifted by $\SI{0.01}{\nano\volt\squared\per\hertz}$ relative to each other, starting with the dark blue data.

\FloatBarrier

\section{Used fit parameters of the theoretical models}
From the fit to the off-resonance noise power spectral density $S_{\rm{n}}$ (see Eq.~(1) in the main manuscript), we obtain the additional constant offset $S_{\rm{off}}=\SI{0.006(0.001)}{\nano\volt\squared\per\hertz}$. 
By using the in-plane Kittel function to fit the broadband FMR data and assuming a vanishing uniaxial magnetic anisotropy field, we obtain a saturation magnetization $M_{\rm{s}} = \SI{134.9(0.2)}{\kilo\ampere\per\meter}$ and a gyromagnetic ratio $\gamma/2\pi = \SI{28.30(0.01)}{\giga\hertz\per\tesla}$ for the YIG thin film (cf.~\cite{Kalarickal2006}).
For fitting the complex scattering parameter $S_{21}(\omega)$ as well as the noise power spectral density $S_{\rm{n}}(\omega)$ (cf. Fig.~4 in the main text), we keep $S_{\rm{off}}$, $M_{\rm{s}}$, $\gamma$ and the external magnetic field $\mu_0 H_0 = \SI{170.2}{\milli\tesla}$ constant.
Please note that in the expression for $S_{\rm{n}}(\omega)$ (cf. Eq.~(3) of the main text) $\left| \exp(i \alpha)\exp(-i \tau \omega)\right|^2 = 1$  is equal to $1$.
Table~\ref{tab:params} shows the numerical parameters obtained from fitting $S_{21}(\omega)$ and $S_{\rm{n}}(\omega)$ to Eq.~(2) and (3) of the main text.

\begin{table}[h]
    \caption{Summary of the numerical parameters obtained from fitting $S_{21}(\omega)$ and $S_{\rm{n}}(\omega)$.}
    \label{tab:params}
    \begin{tabular}{@{}cccccccc@{}}
        \toprule
          & $A$ & $\eta$ & $\phi$ & $\alpha$ & $\tau$ & $\Delta\omega/2\pi$ & $\omega_{\rm{res}}/2\pi$ \\
          & & & (rad) & (rad) & (ns) & (MHz) & (GHz) \\
        \midrule
        $S_{21}(\omega)$ & \num{0.4528(0.0002)} & \num{0.001285(0.000004)} & \num{0.058(0.003)} & \num{2.85(0.05)} & \num{10.566(0.001)} & \num{11.43(0.04)} & \num{6.99665(0.00002)} \\
        $S_{n}(\omega)$ & \num{0.479(0.001)} & \num{0.00135(0.00005)} & \num{0.058} (fixed) & -- & -- & \num{11.1(0.3)} & \num{6.99674(0.00009)} \\
        \bottomrule
    \end{tabular}
\end{table}


\FloatBarrier

\bibliography{references}